\documentclass[10pt,letterpaper,onecolumn]{article}
\usepackage{amsmath,amssymb,graphicx,cite,url,float,dsfont}
\usepackage[top=0.85in,bottom=0.85in,left=1.41in,right=1.41in,footskip=0.45in]{geometry}
\begin{document}
\begin{flushleft}
{\Large\textbf\newline{Single-pixel imaging based on weight sort of the Hadamard basis}}

\bigskip
Wen-Kai Yu\textsuperscript{1,2,*},
Chong Cao\textsuperscript{1,2},
Ying Yang\textsuperscript{1,2},
Ning Wei\textsuperscript{1,2},
Shuo-Fei Wang\textsuperscript{1,2},
Chen-Xi Zhu\textsuperscript{1,2}
\\
\bigskip
$^1$ Center for Quantum Technology Research, School of Physics, Beijing Institute of Technology, Beijing 100081, China\\
$^2$ Key Laboratory of Advanced Optoelectronic Quantum Architecture and Measurement of Ministry of Education, School of Physics, Beijing Institute of Technology, Beijing 100081, China\\
* yuwenkai@bit.edu.cn

\end{flushleft}

\date{}


\noindent\textbf{Abstract:} Single-pixel imaging (SPI) is very popular in subsampling applications, but the random measurement matrices it typically uses will lead to measurement blindness as well as difficulties in calculation and storage, and will also limit the further reduction in sampling rate. The deterministic Hadamard basis has become an alternative choice due to its orthogonality and structural characteristics. There is evidence that sorting the Hadamard basis is beneficial to further reduce the sampling rate, thus many orderings have emerged, but their relations remain unclear and lack a unified theory. Given this, here we specially propose a concept named selection history, which can record the Hadamard spatial folding process, and build a model based on it to reveal the formation mechanisms of different orderings and to deduce the mutual conversion relationship among them. Then, a weight ordering of the Hadamard basis is proposed. Both numerical simulation and experimental results have demonstrated that with this weight sort technique, the sampling rate, reconstruction time and matrix memory consumption are greatly reduced in comparison to traditional sorting methods. Therefore, we believe that this method may pave the way for real-time single-pixel imaging.

\section{Introduction}
Single-pixel imaging (SPI) \cite{Edgar2019,Gibson2020} provides an indirect measurement imaging modality, where only a single-pixel detector is used to collect the total light intensities that correspond to the linear projections between the target and spatially modulated patterns. As we know, the use of single-pixel detector provides scalable designs in various wavelengths (especially in invisible ranges), low-light illumination, hyperspectral imaging \cite{Studer2012,Bian2016} applications, and also offers possibilities for imaging through scattering media. In another aspect, the SPI gives a resource-efficient alternative to traditional pixelated array detectors, allowing one to image the object without raster scanning. As one of the SPI methods, computational ghost imaging \cite{Shapiro2008,Bromberg2009} can retrieve the object images by simply calculating the intensity correlation between modulated patterns and single-pixel measurements, but with a relatively poor image quality, and it suffers from oversampling, long time acquisition and large storage. This scheme will not be too favored in image acquisition applications (e.g., biological imaging, x-ray medical imaging) because the oversampling in these scenarios will harm the targets to be detected. To solve these problems, one can apply compressed sensing (CS) \cite{Donoho2006,Tao2006,Candes2008} for sub-sampling image acquisition and high-quality reconstruction. In CS, it suggests that the natural image signal is generally compressive (containing a large amount of redundant information) and has a sparse representation in some domain \cite{Donoho2006}. Thus, by exploiting the image sparsity, we can recover the image signal from much fewer measurements than those prescribed by the Nyquist-Shannon sampling criterion \cite{Duarte2008}.

In the SPI schemes, a number of time-varying spatial patterns are displayed on a spatial light modulator (SLM). Then, the image details will be built up according to multiple sequential measurements and modulated patterns, and the resolution of the recovered image depends on the resolution of modulated patterns. Therefore, the modulated patterns are crucial for image reconstruction. The design of measurement matrix has become the main challenge of SPI. A variety of measurement matrices have been studied, such as random matrix, Bernoulli matrix, Poisson matrix, Gaussian matrix \cite{FuliLi2019}, etc. However, the entries of these matrices are generally independent of each other, without considering the structural characteristics of the object and thus resulting in blind sampling. This undoubtedly limits further reduction of the sampling rate. Lately, the measurement matrices trained by deep convolutional neural networks have been reported \cite{Lyu2017,Higham2018,Shimobaba2018,Situ2019} to be beneficial to acquire lower sampling rates, but the physics behind them is still unknown. Besides machine learning, it is widely noted that the optimized ordering of deterministic structural Hadamard basis can generate better qualitative images compared with the aforementioned measurement matrices \cite{Vaz2020}.

Several ordering techniques \cite{Sun2017,YuOR2019,YuCC2019,XYu2020,HWu2020} have been proposed successively to study the effect of the deterministic pattern construction on reconstruction performance. In ``Russian dolls'' (RD) ordering \cite{Sun2017}, the Hadamard basis is catalogued into four quarters, the first quarter is further split into four new quarters, and so forth. Note that a high-order Hadamard matrix contains a scaled version of lower-order Hadamard matrix (scaled by a factor 2), the Hadamard basis will be reordered such that the first half consists of lower-order Hadamard basis and the first quarter is composed of the next lower-order basis. In each layer, the third quarter will be put into the transpose of the second quarter, and the fourth quarter will be put into the rest. After that, the basis patterns within each quarter are rearranged in a descending order according to the speckle coherence area of each pattern. In ``origami'' (OR) ordering \cite{YuOR2019}, although the patterns are generated by symmetric reverse folding, axial symmetry and partial pattern order adjustment steps, they turn out to be a new reordered sequence of the Hadamard basis. In ``cake-cutting'' (CC) ordering \cite{YuCC2019}, the Hadamard basis is sorted by increasing the number of connected domains in each pattern. The RD order can acquire a sampling ratio as low as 6\%, but it presents a sawtooth descent in the relative error of recovered images with the increase of the sampling rate and is sensitive to noise. The OR and CC orders are capable of retrieving images of large pixel-size with super sub-Nyquist sampling ratio even below 0.2\% and outperform the RD order. Since there are specific mathematical laws between the CC order and classic sequency (SE) order (also known as Walsh order), the CC order can be quickly obtained, which makes it possible to call the row matrix multiplication operation for quick reconstruction. As for the RD and OR orders, they have no mathematical rules with the sequency order and lack a quick generation method. It is worth mentioning that apart from the sequency order, the Hadamard basis also has other conventional orders, e.g., the natural (NA) order and dyadic order. Besides, we also need to note that all these sorts are mainly used for the rows of the Hadamard matrix, which will be reshaped into the Hadamard basis patterns.

In this paper, we will present a study on how the Hadamard basis orderings affect the reconstruction quality. A mathematical model is built on the basis of a concept we call selection history to reveal the spatial folding mechanism of the Hadamard basis. Based on this model, we further develop a weight (WH) sort of the Hadamard basis, which can realize super sub-Nyquist sampling. The performance of this proposed approach will be demonstrated by both numerical simulation and proof-of-principle SPI experiments. Here, we will use a reconstruction algorithm named as ``total variation minimization by augmented Lagrangian and alternating direction'' (TVAL3) \cite{CBLi2010} to solve the underdetermined problem (where the number of equations or measurements is much less than the number of unknowns). Six different orderings of the Hadamard basis (including NA, SE, RD, OR, CC and WH orderings) are investigated for comparisons. The first two orderings of the Hadamard basis are chosen because they are classic and representative, while the RD, OR and CC are selected for the fact that they stand for new variant sorts that are very close to the weight sort proposed in this work.

\section{Theory}
\subsection{Theory of the SPI}
The essence of the SPI is to change the way of acquiring spatial information of the target from pixel-wise array detection to single-pixel compressive sampling. Generally, the SPI can be divided into an encoding process and a decoding process. In the encoding process, the object image will be encoded by modulated patterns into a sequence of single-pixel measured values; while in the decoding process, the spatial distribution of the target will be recovered according to recorded total intensities and modulated patterns. The mathematical measurement model can be written as $y=Af+e$, where $y$ denotes the single-pixel measurement vector, $A$ stands for the measurement matrix, $f$ represents the column vector which is flattened from the original image of $p\times q$ pixels, and $e$ is the stochastic measurement noise. As mentioned above, the measurement matrix can be a Hadamard matrix $H$ \cite{Decker1970}, which is a symmetric $N\times N$ square (full-rank) matrix ($H=H^T$) consisting of $\pm1$. $N=p\times q=2^k$ is a power of 2, $k$ is a positive integer. There is $HH^T=NI_N$, i.e., the inverse of $H$ divided by $\sqrt{N}$ is itself, where $I_N$ is an $N\times N$ identity matrix, and $T$ stands for the transpose symbol. For modulation, each row (basis) of the Hadamard matrix can be reshaped to a pattern of $p\times q$ pixels, which we call the Hadamard basis pattern.

Generally, a natural image is compressible and can be sparsely represented in some basis $\Psi$, i.e., $f=\Psi f'$. The top $K\ll N$ largest sparse representation coefficients (most of them are the low-frequency components) are sufficient to represent the main image information, and the rest (most of them correspond to the high-frequency components) can be set to 0. We can define a sparsity as the ratio of the number of non-zeros (large-values) among the sparse representation coefficients to the total number of image pixels, i.e., $\alpha=K/N$. It is asserted by the CS theory that the object image can be reconstructed with high-quality from much fewer measurements by exploiting the sparsity of the object. The number of measurements is $M=O(K\cdot\log(N/K))<N$, which is generally greater than or equal to the sparsity $K$. It means that the measurement matrix $A$ should be of $M\times N$ and the sampling ratio can be defined as $M/N$. For random sampling, we can directly generate a pure random or pseudo-random matrix of $M\times N$, and we can also randomly permute both columns and rows of the Hadamard matrix and then select $M$ rows to acquire a totally random binary matrix of $M\times N$ \cite{CBLi2010}. While in the cases of using optimized Hadamard ordering, only the rows of the Hadamard matrix are rearranged. A straightforward approach is to randomly rearrange the rows of the Hadamard matrix, but it will not acquire as good performance as a pure random measurement matrix \cite{YuCC2019}.

\subsection{Selection history for the Hadamard matrix}
\begin{figure}[htbp]
\centering\includegraphics[width=\textwidth]{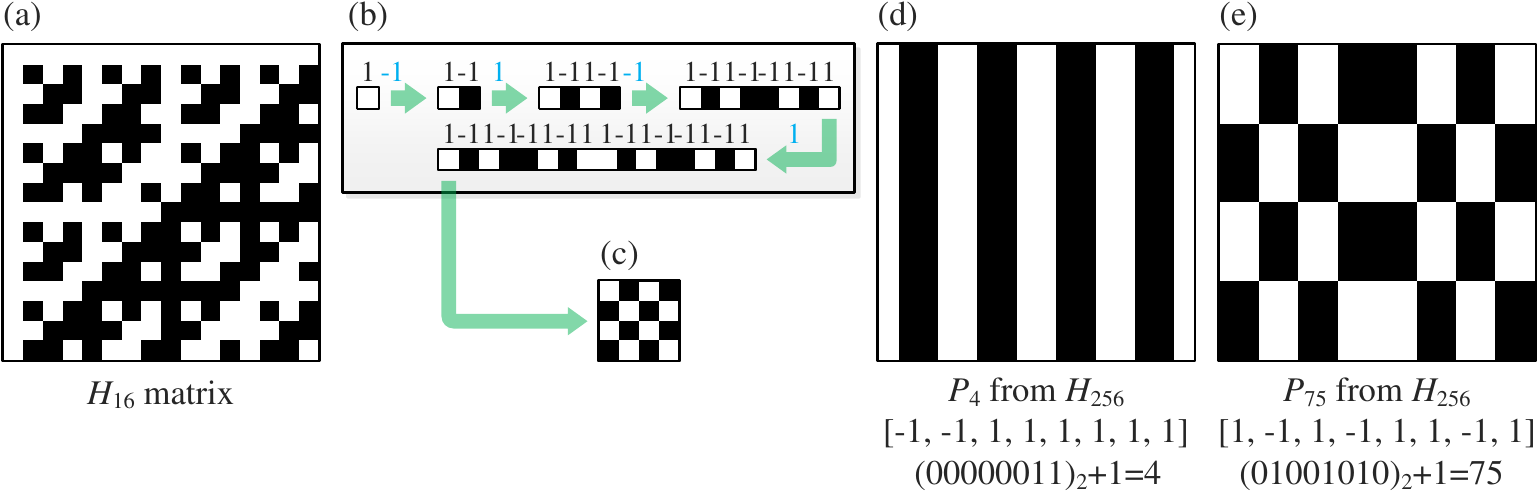}
\caption{Selection history and pattern formation process. (a) is a $16\times16$ Hadamard matrix $H_{16}$; (b) is the selection history of sixth row of the matrix $H_{16}$; (c) is the Hadamard basis pattern reshaped from the sixth row of the matrix $H_{16}$; (d)--(e) are two examples of 8-bit binary numbers and the $256\times256$ modulation patterns generated by these two binary numbers.}
\label{fig1}
\end{figure}

The naturally ordered Hadamard matrix follows the following recursive formula:
\begin{equation}
H_{2^k}=
\left[
\begin{array}{cc}
H_{2^{k-1}}&H_{2^{k-1}}\\
H_{2^{k-1}}&-H_{2^{k-1}}
\end{array}
\right]
=H_2\otimes H_{2^{k-1}},
\end{equation}
where $2^k\in[2,N]$ is the order of the Hadamard matrix, $H_1=[1]$, $H_2=\left[{\begin{array}{*{20}{c}}
1&1\\
1&-1
\end{array}}\right]$, and $\otimes$ represents the Kronecker product. The following properties can be derived: 1) the length of each row of the matrix $H_{2^k}$ is twice that of $H_{2^{k-1}}$; 2) the first half entries of the second half of the rows (i.e., the last $2^{k-1}$ rows) of the matrix $H_{2^k}$ are the same as those of the first half of the rows (i.e., the first $2^{k-1}$ rows) of the matrix $H_{2^k}$; 3) for the first half of the rows (i.e., the first $2^{k-1}$ rows) of matrix $H_{2^k}$, the second half entries of each row are the same as the first half entries, we call the selection value of the second half entries relative to the first half entries as 1; 4) for the second half of the rows (i.e., the last $2^{k-1}$ rows) of matrix $H_{2^k}$, the second half elements of each row are exactly the opposite of the first half elements, we call the selection value of the second half elements relative to the first half elements as -1. According to these properties, the iterative process from $H_{2^{k-1}}$ to $H_{2^k}$ is to set the selection values with respect to $H_{2^{k-1}}$ as 1 and -1, respectively, and then concatenate these two generated matrices up and down. Take $H_2$ as an example, it is to set the selection values with respect to $H_1$ as 1 and -1, respectively. Following this rule, the matrix $H_{2^k}$ is the result of selecting $k$ sets in total (each set contains one option 1 and one option -1) with respect to the matrix from $H_1$ to $H_{2^{k-1}}$. The above selection operations are all for rows, and can also apply to selection operations on columns. Arranging these selection values in order of selection generates a set we call selection history. We mark 1s in these selection values as 0s, and -1s as 1s, and then arrange these values in order from the lowest bit to the highest bit to form a binary number. The binary number plus 1 is the serial number corresponding to this selection history, i.e., the row number or the column number. Plus 1 is because the serial number of the pattern stats from 1 rather than 0. Take the sixth row of $16\times16$ Hadamard matrix $H_{16}$ (see Fig.~\ref{fig1}(a)) for example, the selection history is [-1, 1, -1, 1], the sequence after being marked is [1, 0, 1, 0], the formed binary number is $(0101)_2$, and the serial number is $(0101)_2+1=5+1=6$. From this example, we can clearly see that the generation of the Hadamard matrix is a process of spatial folding. The above rule can be easily proved via mathematical induction.

\noindent\textbf{\textit{Proof.}} Suppose the binary number of the $i$th row of the Hadamard matrix $H_{2^{k-1}}$ is $(B^{(k-1)}_i)_2$, there will be an equation $i=(B^{(k-1)}_i)_2+1$. Assume the binary number of the $j$th row of the Hadmard matrix $H_{2^k}$ is $(B^{(k)}_j)_2$, then we have:

1) when $1\leq j\leq2^{k-1}$, there is $j=i$, and the $k$th selection value should be 1 according to the properties of the naturally ordered Hadamard matrix, we compute $(B^{(k)}_j)_2+1=(0B^{(k-1)}_i)_2+1=0+(B^{(k-1)}_i)_2+1=i$, so $j=(B^{(k)}_j)_2+1$;

2) when $2^{k-1}<j\leq2^k$, there is $j=i+2^{k-1}$, and the $k$th selection value should be -1 according to the properties of the naturally ordered Hadamard matrix, we calculate $(B^{(k)}_j)_2+1=(1B^{(k-1)}_i)_2+1=2^{k-1}+(B^{(k-1)}_i)_2+1=i+2^{k-1}$, so $j=(B^{(k)}_j)_2+1$.

Therefore, we can deduce that $j=(B^{(k)}_j)_2+1$. $\blacksquare$

Then, each row of the Hadamard matrix can be reshaped into a Hadamard basis pattern to be modulated, by sorting in row-major order (a general practice). Figure~\ref{fig1}(c) gives an example of the Hadamard basis pattern that is reshaped from the sixth row of $H_{16}$.

Conversely, when we know $N$, we can also deduce the binary sequence $(B_i)_2=i-1$ according to the serial number $i$. Taking a $256\times256$ Hadamard matrix for example, $N=256=2^8$, there will be 8 sets of selection for each row's generation, and the Hadamard basis patterns are all of $16\times16$. If the chosen row (serial) numbers are 4 and 75, the corresponding binary numbers are $(00000011)_2$ and $(01001010)_2$, the selection histories can be easily obtained as [-1, -1, 1, 1, 1, 1, 1, 1] and [1, -1, 1, -1, 1, 1, -1, 1]. According to these selection histories, we can quickly generate the corresponding rows of the Hadamard matrix, which will be then reshaped in row-major order to form the Hadamard basis patterns, as shown in Figs.~\ref{fig1}(d)--\ref{fig1}(e).

In Fig.~\ref{fig1}(d), we generated the pattern of $16\times16$ pixels from the fourth row of $H_{256}$ as an example, whose rows are exactly the same. This pattern actually selected 1 for the first row vector four times. Given this, we began to think about whether we can directly obtain the $i$th Hadamard basis pattern from the selection history, but without the need for the matrix reshaping operation. We found that the selection history can be divided into two parts: the first half correspond to the row (or column) entries' folding operations starting from the first entry, while the second half correspond to the row (or column) folding operations starting from the newly generated first row (or column) of the Hadamard basis pattern. Each selection will double the length of the row (or the column) vector. Also taking the sixth row of the matrix $H_{16}$ as an example, its selection history is [-1, 1, -1, 1], the first half selection values are used to build the first row of the sixth Hadamard basis pattern (by copying and inverting operations on cells), and then the second half selection values are applied to generate the rest rows of the sixth Hadamard basis pattern according to this first row just formed. If the serial number picked out is the column number, the selection history should also be for this column, then we can directly generate the Hadamard basis pattern with respect to this column number in the same way. We call a selection value of 1 positive folding and a selection value of -1 negative folding. Hence, the selection history can also be called the folding history.

\subsection{Connected domain}
Now, we have known our selection history can be used to generate the two-dimensional (2D) Hadamard basis pattern by positive-negative folding. With the patterns, a concept named connected domain (region) \cite{YuCC2019} needs to be defined here as a topological pixel region that consists of adjacent pixels (being up and down, left and right connected) with the same pixel value and it cannot be split into two or more disjoint nonempty subsets. The number of connected domains in a 2D Hadamard basis pattern is our focus. But before that, we first need to use the selection history to calculate the number of one-dimensional (1D) connected domains in a 1D vector (one row or one column).

\subsubsection{Calculation for the number of 1D connected domains in a row or column via selection history}
\begin{figure}[htbp]
\centering\includegraphics[width=0.85\textwidth]{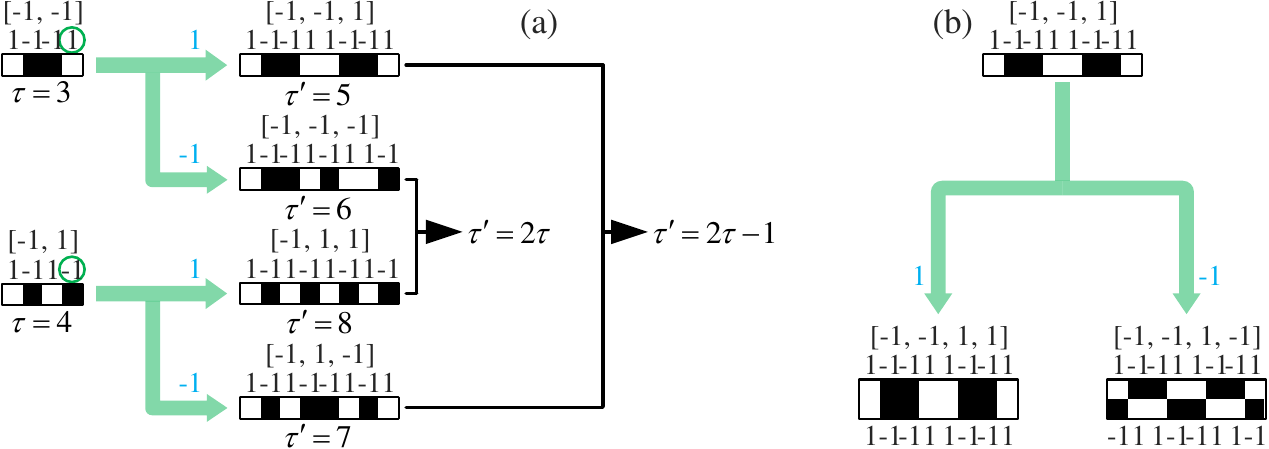}
\caption{Number of (a) 1D and (b) 2D connected domains calculated by selection history.}
\label{fig2}
\end{figure}

One row in the matrix $H_{2^{k-1}}$ can be denoted by $h_{2^{k-1}}=[x_1,x_2,\cdots,x_{2^{k-1}}]$, its number of 1D connected domains $\tau$ is the number of sign (pixel value) changes in $h_{2^{k-1}}$ plus 1. If the next selection value for this row is 1, we will get $h_{2^k}=[h_{2^{k-1}},h_{2^{k-1}}]=[x_1,x_2,\cdots,x_{2^{k-1}},x_1,x_2,\cdots,x_{2^{k-1}}]$. We just need to see whether there is a sign change at the junction $[x_{2^{k-1}},x_1]$ in the middle. If there is a sign change in the middle, then the number of 1D connected domains in $h_{2^k}$ should be $2\tau$; otherwise it is $2\tau-1$. Similarly, if the next selection value for the row $h_{2^{k-1}}$ is -1, the conclusions are the same. Here, we give two examples as shown in Fig.~\ref{fig2}(a), the first original row has the same first and last signs (pixel values), and the second original row has different first and last signs (pixel values). In the first case, the original number of 1D connected domains $\tau$ is 3, when the next selection value is 1, the number of 1D connected domains of the pattern after being flipped $\tau'=2\times3-1=5$; and when the next selection value is -1, $\tau'=2\times3=6$. While in the second case, $\tau=4$, when the next selection value is 1, $\tau'=2\times4=8$; and when the next selection value is -1, $\tau'=2\times4-1=7$. This is consistent with the previous theoretical analysis.

The first row and the first column in the naturally ordered Hadamard matrix are all ones, so the starting pixel value in each row or column is 1. Therefore, the sign (pixel value) of $x_{2^{k-1}}$ is just the product of the previous $k-1$ selection values. We can generalize this deduction to more general case, i.e., the sign (pixel value) of $x_{2^{k-1}}$ is the product of the previous $k-1$ selection values and the sign (pixel value) of the first entry $x_1$. Here, the sign (pixel value) of $x_1$ can be 1 or -1. This conclusion also applies to each column. Based on this conclusion, we can update the value of $\tau'$ by comparing the next selection value with the product of current selection history and the sign of the first element. If they are the same, then $\tau'=2\tau-1$, otherwise $\tau'=2\tau$. According to this method, we can calculate the number of 1D connected domains of $h_{2^k}$ from the starting element.

\subsubsection{Calculation for the number of 2D connected domains in the Hadamard basis pattern via selection history}
As mentioned above, the first row (column) of one Hadamard basis pattern $P_i$ can be generated according to the first entry and the first half of the selection history, and the rest rows (columns) can be formed according to the first row (column) just generated and the second half of the selection history. Therefore, the number of 2D connected domains of this Hadamard basis pattern is equal to the product of the first row's and the first column's numbers of 1D connected domains. Thereby, the number of 2D connected domains in one Hadamard basis pattern can be quickly calculated by our selection history. As shown in Fig.~\ref{fig2}(b), the initial number of 1D connected domains of the row vector is 5, when the next selection value is 1, the first column's number of 1D connected domains is 1, then the number of 2D connected domains is $\tau'=5\times1=5$; and when the next selection value is -1, the first column's number of 1D connected domains is 2, we have $\tau'=5\times2=10$.

It is worth mentioning that there are also general algorithms for calculating the number of 2D connected domains of images, such as depth first search algorithm or using MATLAB built-in bwlabel function. However, the computational complexity of these general algorithms is too high. The calculation method based on the selection history can dramatically reduce the computational complexity.

\subsection{Selection-history-based theoretical explanations for different\\Hadamard basis orderings' formation mechanisms}
Next, we will use our selection history concept to explain a variety of mainstream ordering strategies of the Hadamard basis, including SE, RD, OR and CC orderings.

We have already established calculation methods for the number of 1D or 2D connected domains by using the selection history. As we all know, the SE order is an increasing ordering of the number of sign changes in each row of the naturally ordered Hadamard matrix, and the number of sign changes is equivalent to the number of 1D connected domains. Thus, we can realize the mutual conversion between the SE ordering and RA ordering according to the calculation method based on selection history introduced earlier. Similarly, the CC order is an increasing ordering of the numbers of 2D connected domains of the Hadamard basis patterns, so the mutual conversion between the CC ordering and RA ordering can be done by using the idea of selection history. It is worth mentioning that the mutual conversion between the CC ordering and SE ordering has already realized in our previous work \cite{YuCC2019}.

As mentioned before, in the RD method \cite{Sun2017}, the Hadamard basis patterns need to be catalogued into four quarters, the patterns in each of which will be then rearranged in a descending order of the speckle coherence area (or an increasing order of the number of 2D connected regions) of each pattern. The first step requires the Hadamard matrix $H_{2^k}$ are reordered such that the first half rows are the rows of a Hadamard matrix $H_{2^{k-1}}$ (scaled by a factor 2), the first quarter rows are the rows of a Hadamard matrix $H_{2^{k-2}}$ (scaled by a factor 4), the first eighth rows are the rows of a Hadamard matrix $H_{2^{k-3}}$ (scaled by a factor 8), and so on. Following this rule, the rows in the first and the second quarter are fixed. Since the Hadamard matrix has symmetry, the transpose of the second quarter can always be found in the rows of the second half of the matrix $H_{2^k}$. Thus, the third quarter is ordered as the transpose of the second quarter, and the fourth quarter will be put into the rest. This means that the patterns in the second quarter and the third quarter are row- or column-stretched patterns, respectively. According to the theory of selection history, the pixel dimensions of the rows (or columns) of these patterns are doubled, and the pattern image is only stretched without changing the number of 2D connected domains, we only need to add a 1 before the first selection value of the current selection history for the first row (or the first column) of the pattern to generate a new selection history. Without loss of generality, we assume that the patterns are all reshaped from rows of the Hadamard matrix in row-major order, and that the selection history of each low-order Hadamard basis pattern $P_{2^{k-2}}$ is [$a_1$, $a_2$, $\cdots$, $a_{(k-2)/2}$, $b_1$, $b_2$, $\cdots$, $b_{(k-2)/2}$]. These patterns are actually reshaped from rows of $H_{2^{k-2}}$ and need to be scaled up by a factor 4, thus the selection history for each pattern in the first quarter should be [1, $a_1$, $a_2$, $\cdots$, $a_{(k-2)/2}$, 1, $b_1$, $b_2$, $\cdots$, $b_{(k-2)/2}$]. Note that the second and third quarters are interchangeable and are transpose of each other, the selection histories for the patterns in the second and third quarters should be either [1, $a_1$, $a_2$, $\cdots$, $a_{(k-2)/2}$, -1, $b_1$, $b_2$, $\cdots$, $b_{(k-2)/2}$] or [-1, $a_1$, $a_2$, $\cdots$, $a_{(k-2)/2}$, 1, $b_1$, $b_2$, $\cdots$, $b_{(k-2)/2}$]. Then, the selection history for each pattern in the fourth quarter has to be [-1, $a_1$, $a_2$, $\cdots$, $a_{(k-2)/2}$, -1, $b_1$, $b_2$, $\cdots$, $b_{(k-2)/2}$]. The iteration in each layer follows the same rules. After that, we calculate the number of 2D connected domains in patterns of each quarter via selection history and rearrange them to form the complete Hadamard basis patterns of $H_{2^k}$ in the RD order.

In the OR method, the Hadamard basis patterns are catalogued into $N/4$ groups, each of which contains four patterns. Starting from a pattern of all ones (whose selection history is [1, 1]), the second and third patterns in the first group are obtained by inversely folding the first pattern up and down, left and right, respectively. The selection histories for these two operations are [1, -1] and [-1, 1] (in row-major order). The fourth pattern in the first group is formed by performing both up-down and left-right reverse folding operations, thus the corresponding history is [-1, -1]. The patterns in the first group need to be scaled up by a factor of $N^2/4$ selection history. That is, we need to add $2\log_4N-1$ ones before the first row's selection value and the first column's selection value, respectively. The first pattern in the $i$th group (also ($4i-3$)th pattern in the complete sequency) is formed by selecting [1, -1] on the $i$th pattern in the current complete sequence. For each group, the scaled factor will be adjusted accordingly. At last, adjust the partial pattern order so that the number of 2D connected domains in each pattern of $i$th group is in an ascending order, also using the calculation method based on selection history. Therefore, the OR ordering can also be deduced from our selection history.

The above four sorting methods all involve the calculation and sorting of the number of connected domains. Compared with the famous SE order, the CC order changes the sorting of the number of 1D connected domains to that of 2D connected domains, the former applies to 1D signals and the latter applies to 2D images, both are of great scientific importance. In particular, the evolution from the RD to the OR is essentially changing the division from the outer to the inner, thereby eliminating the jagged oscillating growth of image quality with the increase of sampling rate.

\subsection{Weight sort of the Hadamard basis}
Next, let us take a closer look at the selection history with respect to a Hadamard basis pattern. Also take the Hadamard matrix $H_{2^k}$ as an example, and without loss of generality we assume the pattern is reshaped in row-major order, there will be a total of $k$ selection values in the selection history of this pattern, where the first $k/2$ selection values correspond to the horizontal folding of the starting seed element (taking either 1 or -1, in the natural order it is 1), and the last $k/2$ values correspond to the vertical folding of the seed row just generated. Obviously, the first choice value in the first half (i.e., performing the first folding on the seed entry) and the first choice value in the second half (i.e., doing the first folding on the seed row) are the most important for determining the number of 2D connected domains, and the former is more important than the latter (this rule also applies to the column-major order), If they take 1, the number of 2D connected domains in the pattern tends to be smaller with a larger magnitude. Then, the 2nd (i.e., performing the second folding on the first two entries) and the $(k/2+2)$th (i.e., doing the second folding on the first two rows) selection values taking 1 are the second most important; the 3rd and the $(k/2+3)$th selection values taking 1 are the third most important; by analogy, the $(k/2)$th and the $k$th selection values taking 1 are the least important. Among them, the former values are always more important than the latter values. We can therefore sort the importance of the position number of the selection values in the selection history: $1>k/2+1\gg2>k/2+1\gg3>k/2+3\gg\cdots\gg k/2>k$. Since each selection value can be 1 or -1, we might as well use the binary thinking to weight these ranks. For the quantification, the weights for these ranks can be simply set to $2^{k-1}$, $2^{k-2}$, $2^{k-3}$, $\cdots$, $2^2$, $2^1$, $2^0$, and 1s, -1s in these selection values are rewritten as 0, 1.These 0-1 values are arranged in order from the highest bit to the lowest bit to form a binary number. By calculating the weighted sum of bits, we can acquire a serial number. This binary bit arrangement order is exactly opposite to the one in the process of deriving the serial number from the selection history.

\begin{figure}[htbp]
\centering\includegraphics[width=\textwidth]{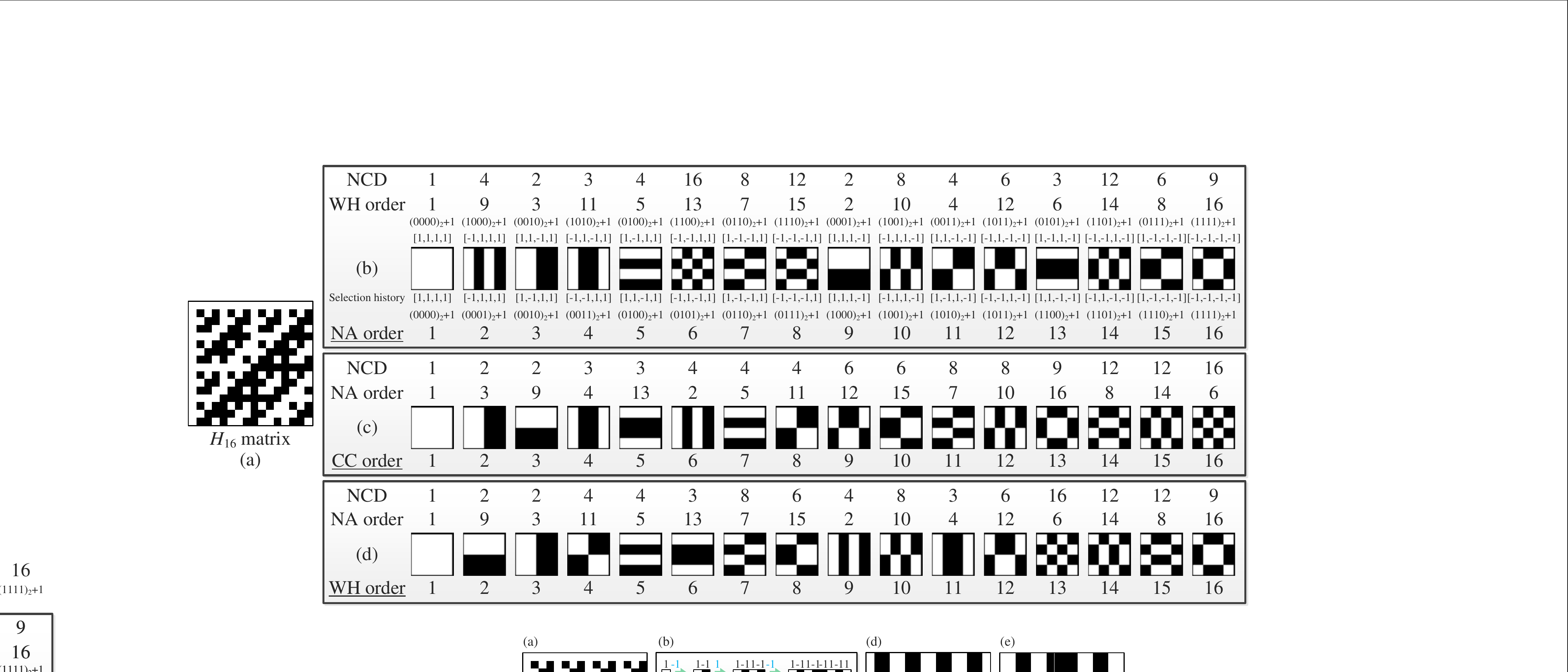}
\caption{Weight ordering formation. (a) is a $16\times16$ Hadamard matrix, (b)--(d) are natural, cake-cutting and weight orderings, respectively. The NCD is short for the number of 2D connected domains, NA, CC and WH orders stand for natural, cake-cutting and weight orderings.}
\label{fig3}
\end{figure}

Here, we provide an example: assume that a pattern extracted from the naturally ordered Hadamard matrix has a selection history [1, -1, -1, 1, 1, 1, -1, -1] (in row-major order), its serial number for the NA order is $(11000110)_2+1=(2^7+2^6+2^2+2^1)+1=199$. Rearranging the selection values according to the weights, we can get [1, 1, -1, 1, -1, -1, 1, -1] and the serial number for the WH order is $(00101101)_2+1=(2^5+2^3+2^2+2^0)+1=46$. According to the above rules, also taking $H_{16}$ as an example (see Fig.~\ref{fig3}(a)), we can get the corresponding selection histories according to the patterns in the natural order (the selection histories can be converted into serial numbers and vice versa), calculate the corresponding numbers of 2D connected domains, and acquire the weight order, as shown in Figs.~\ref{fig3}(b) and \ref{fig3}(c). For comparison, here we also provide the cake-cutting order as a reference.

\begin{figure}[htbp]
\centering\includegraphics[width=\textwidth]{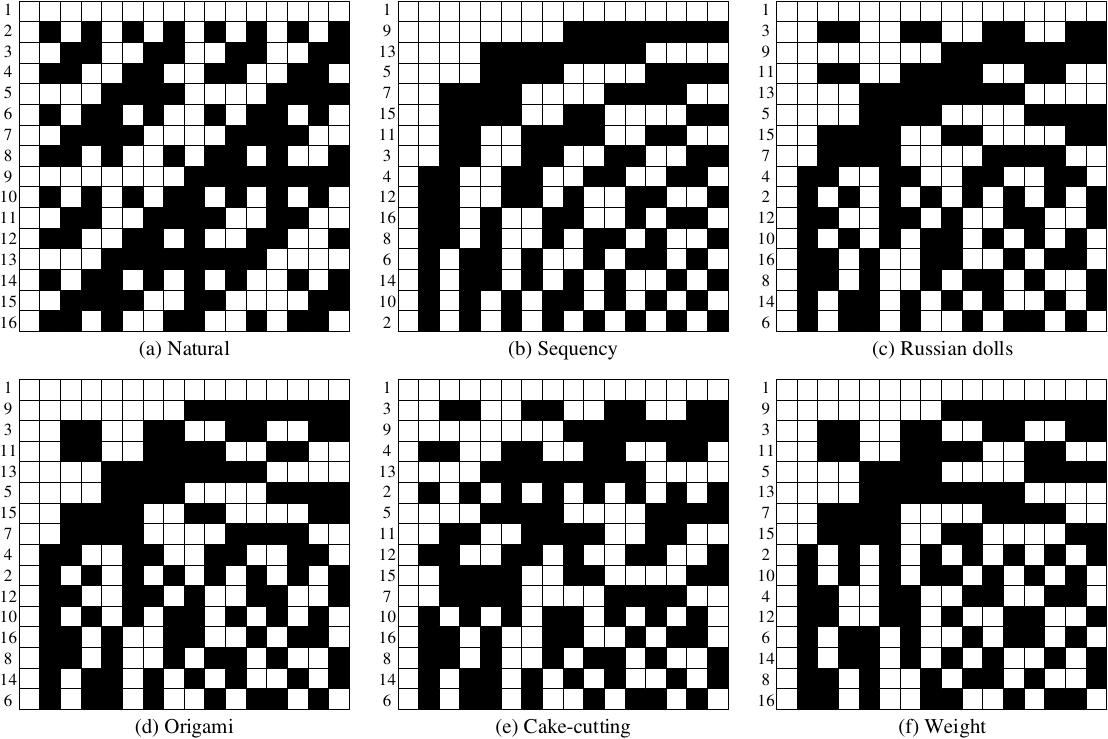}
\caption{Hadamard matrices of $16\times16$ by using (a) natural, (b) sequency (Walsh), (c) Russian dolls, (d) origami, (e) cake-cutting and (f) weight orderings.}
\label{fig4}
\end{figure}

To observe the differences among the NA, SE, RD, OR, CC and WH orders more intuitively, we give the complete Hadamard matrices corresponding to these orders, as shown in Fig.~\ref{fig4}. Without loss of generality, here we also took the Hadamard matrices of $16\times16$ as an example, when the pixel dimension is further increased, their laws and characteristics remain almost unchanged. It can be clearly seen from the sub-figures that the NA-ordered Hadamard matrix has the characteristics of block periodic folding, and the rest matrices have the structural features of radiating from the upper left corner to the lower right corner. The NA-, SE- and WH-ordered Hadamard matrices are all symmetrical, i.e., $H=H^T$. According to the literature \cite{YuOR2019}, the OR and RD orderings have a lot of overlapping serial numbers in low order, but will present a lot of differences in high order. As for the CC order, it is a great promotion of the SE order in the spatial dimension by developing the ascending order of the number of 1D connected domains into the ascending order of the number of 2D connected domains. In addition, the WH-ordered matrix looks similar to the CC-ordered matrix in morphology, but with the use of weights, the Hadamard matrix is changed from a disordered structure to an ordered symmetric structure, which also makes some Hadamard basis patterns with a large number of 2D connected domains be displayed in advance. This operation helps to modulate high-frequency object details at low sampling rates. Therefore, by utilizing the importance of pattern spatial folding, the WH method optimizes the appearance order of the Hadamard basis patterns to a certain extent.

\section{Results}
\subsection{Numerical simulation results}
\begin{figure}[htbp]
\centering\includegraphics[width=\textwidth]{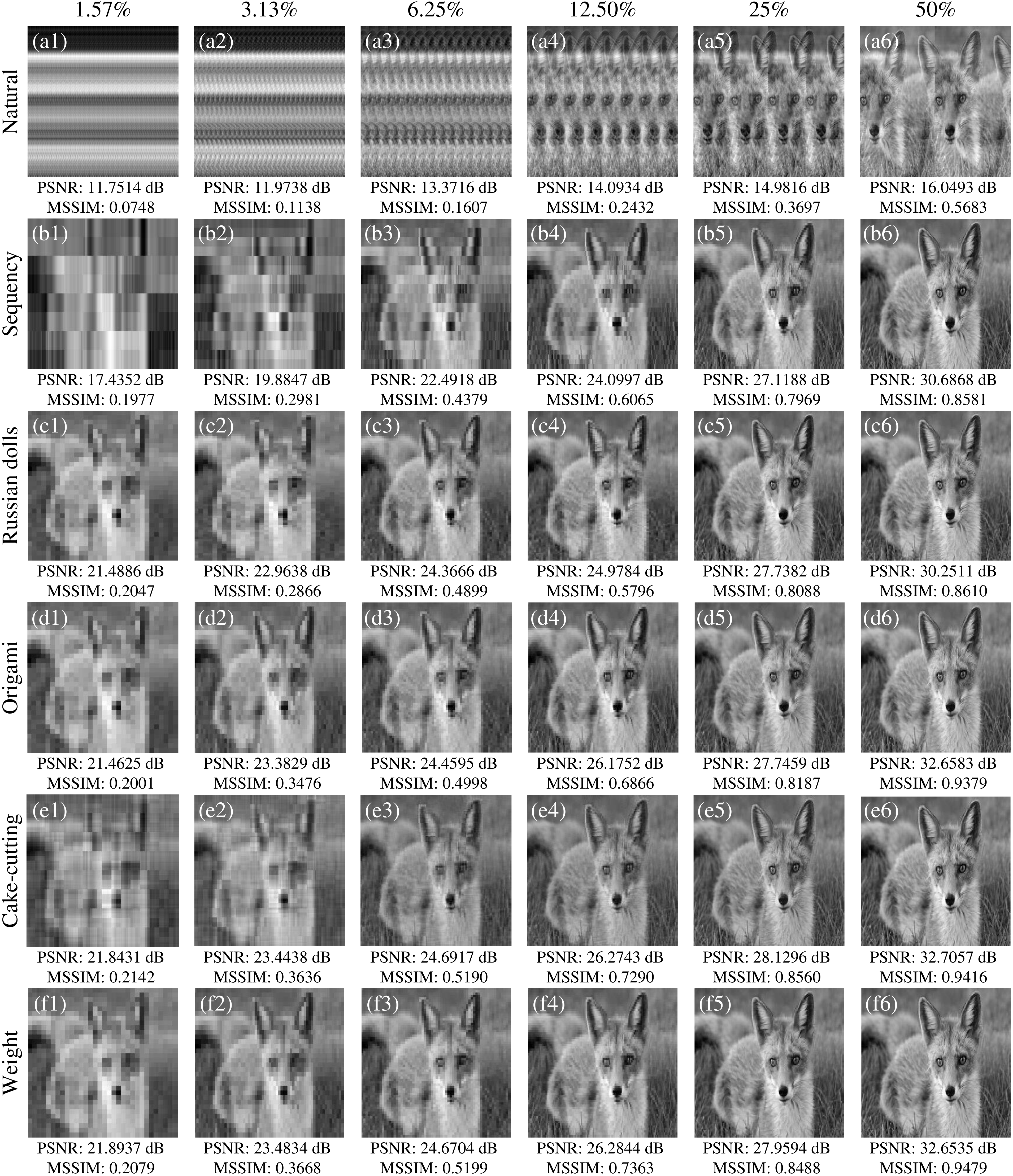}
\caption{Reconstruction for a grayscale image ``fox''. (a1)--(a6), (b1)--(b6), (c1)--(c6), (d1)--(d6), (e1)--(e6) and (f1)--(f6) are the images recovered by natural, sequency, Russian dolls, origami, cake-cutting and weight sorting methods, respectively, all of $256\times256$ pixels.}
\label{fig5}
\end{figure}

Next, we would do some numerical simulation experiments to verify the imaging performance of the proposed WH sorting method and compare it with classical NA, SE, RD, OR, and CC orderings of the Hadamard basis under different sampling ratios. Here, we used the TVAL3 solver \cite{CBLi2010} as the reconstruction algorithm for these methods, and chose the peak signal-to-noise ratio (PSNR) and mean structural similarity (MSSIM) \cite{Simoncelli2004} as a figure of merit to quantitatively measure image quality. The PSNR can be defined as $\textrm{PSNR}=10\log(255^2/\textrm{MSE})$, where $\textrm{MSE}=\frac{1}{pq}\sum\nolimits_{c,d=1}^{p,q}[U_o(c,d)-\tilde U(c,d)]^2$, $U_o$ and $\tilde U$ denote the original (reference) and reconstructed images, respectively. The MSSIM is another full reference metric which is mainly based on the structural information rather than the pixel error between the reference image and recovered image used in the PSNR. Therefore, the MSSIM fills the gap of the PNSR in image content evaluation, and the use of the PSNR in conjunction with the MSSIM can provide a more objective and comprehensive image quality evaluation. Naturally, the larger are the PSNR and MSSIM values, the better is the image quality retrieved. To make a fair comparison, all reconstructed images are normalized a range of $0\sim255$. The MSSIM value ranges from 0 to 1. A grayscale ``fox'' image with complex background is chosen here as the original image. In Fig.~\ref{fig5}, all images are of $256\times256$ pixels, and the Hadamard matrix is of $65536\times65536$, each row of which will be reshaped to a pattern of $256\times256$ pixels, thus $k=16$. We can see from Figs.~\ref{fig5}(a1)--\ref{fig5}(a6) that the recovered images of the NA ordering have a periodic silhouette ghosting, because of the strong structural features of the naturally ordered Hadamard matrix. By comparison, the SE ordering can improve the image quality to a certain extent, but there will still be a certain ghosting in its reconstructed images under low sampling ratios, as shown in Figs.~\ref{fig5}(b1)--\ref{fig5}(b6). Both the RD and OR can effectively eliminate the ghosting (see Figs.~\ref{fig5}(c1)--\ref{fig5}(c6) and \ref{fig5}(d1)--\ref{fig5}(d6)), while the PSNRs and MSSIMs of the latter are higher than those of the former under any sampling ratio with a high probability, showing the natural advantages of grouping and sorting the number of 2D connected domains within a group. As we know, the CC ordering is an extreme strategy which directly rearranges the patterns in an increasing order of the number of 2D connected domains, thus the CC ordering has a better imaging performance (see Figs.~\ref{fig5}(e1)--\ref{fig5}(e6)) than the RD and OR orderings. It can also be seen from Figs.~\ref{fig5}(f1)--\ref{fig5}(f6) that our WH ordering outperforms the CC ordering under most sampling rates, especially when reconstructing image details at ultra-low sampling ratios.

\begin{figure}[htbp]
\centering\includegraphics[width=\textwidth]{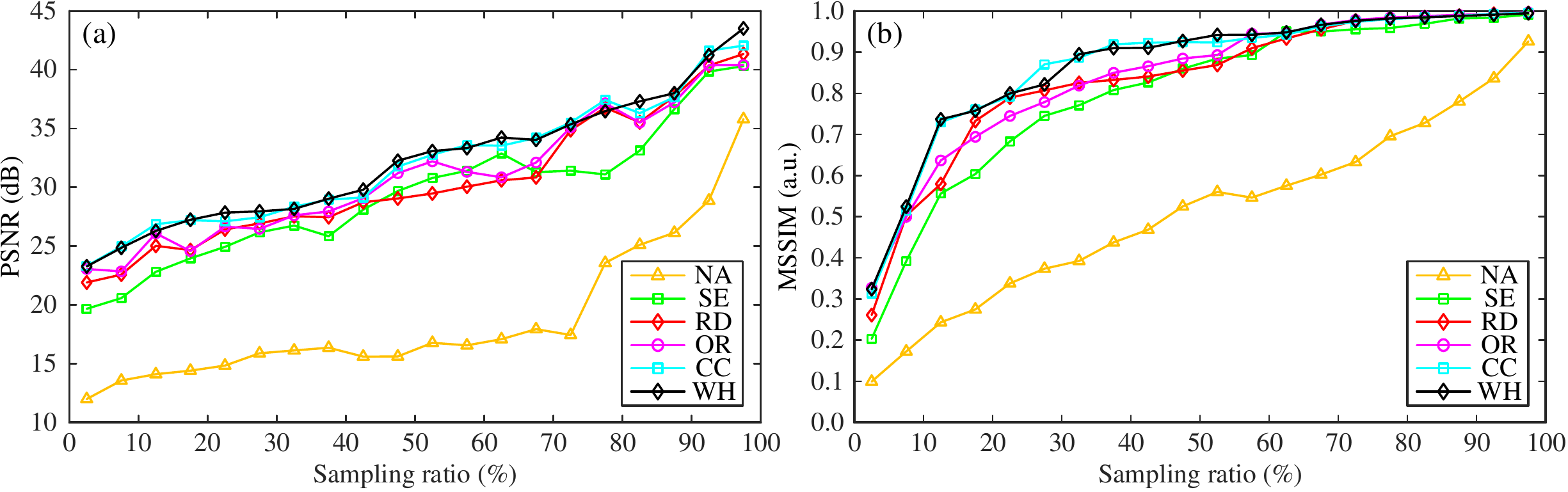}
\caption{Curves of (a) PSNRs and (b) MSSIMs as a function of the sampling ratio, where NA, SE, RD, OR, CC and WH refer to the natural, sequency, Russian dolls, origami, cake-cutting and weight orderings, respectively.}
\label{fig6}
\end{figure}

To give an intuitive comparison of image quality, we drew the curves of PSNRs and MSSIMs as a function of the sampling ratio, as shown in Fig.~\ref{fig6}(a) and \ref{fig6}(b), respectively. From the curves, it can be clearly seen that the image qualities of these methods all improve with the increase of the sampling rate. The imaging performance of the NA ordering is the worst, and it is not suitable for compressive imaging because the Hadamard matrix in the natural order exhibits too obvious structure. The image quality of the SE ordering is better, but is still not ideal, because it mainly sorts the number of 1D connected domains, which is still not very suitable for the 2D image reconstruction. By contrast, the RD and OR methods present better reconstruction performance than the SE ordering, due to the use of the grouping and the sort based on the number of 2D connected domains. And the OR ordering is better than the RD ordering at most sampling rates, because it uses finer grouping. The PSNRs of the CC ordering are significantly higher than those of the previous four orderings, by removing the grouping process. We can also see that our WH sort method exceed the CC ordering approach especially when the sampling ratio is lower than 50\%. This is because the WH method utilizes weights to reintroduce the grouping idea, making the finer patterns more likely to appear in the first half of the ordering.

\begin{figure}[htbp]
\centering\includegraphics[width=\textwidth]{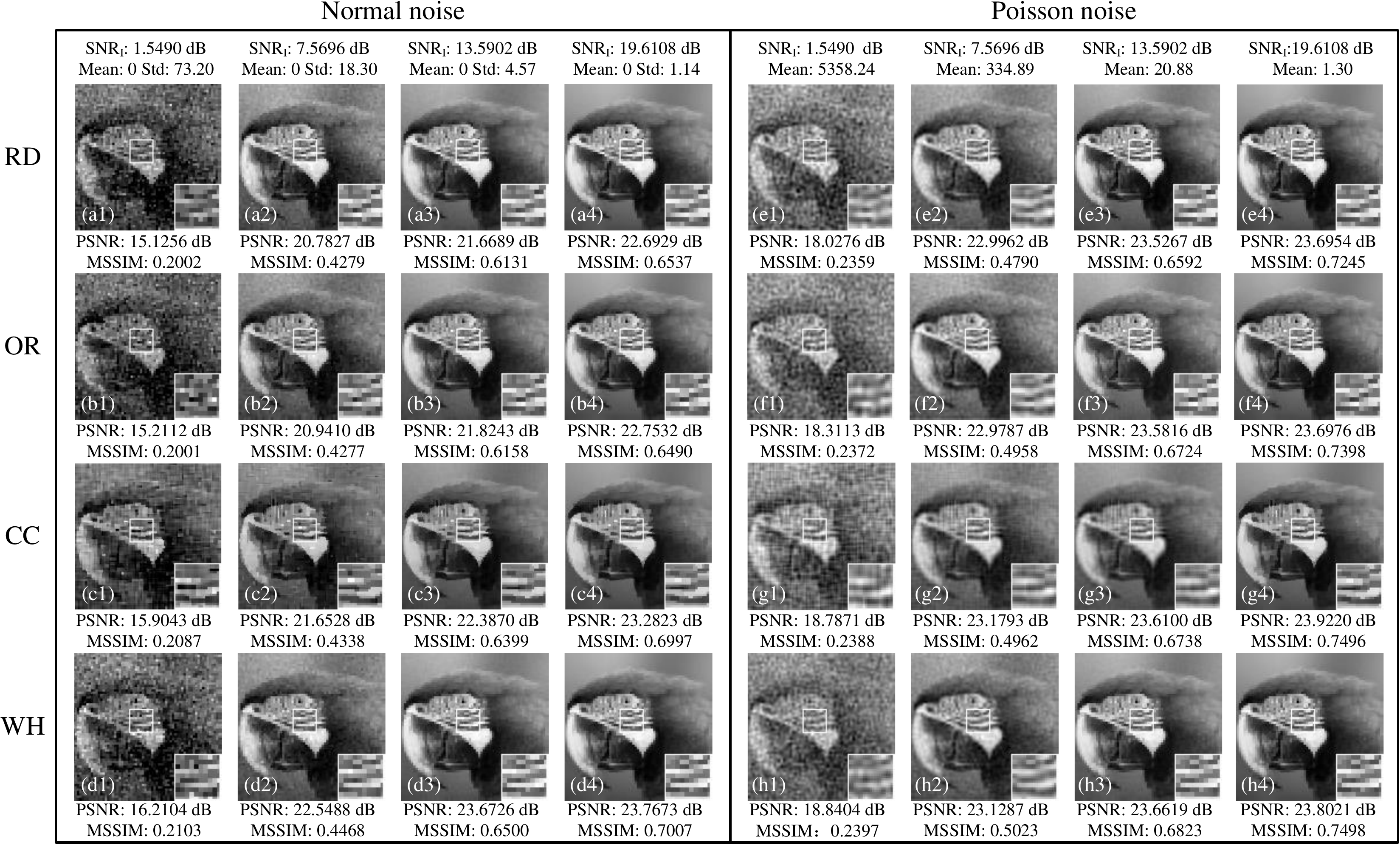}
\caption{Reconstructed images of the RD, OR, CC and WH methods in present of normal and Poisson distributed additive noise. (a1)--(a4), (b1)--(b4), (c1)--(c4) and (d1)--(d4) are the results of the RD, OR, CC and WH methods under normal noise, while (e1)--(e4), (f1)--(f4), (g1)--(g4) and (h1)--(h4) are the images recovered by the RD, OR, CC and WH approaches under Poisson noise. All these images are of $128\times128$ pixels, with a 12.5\% sampling ratio.}
\label{fig7}
\end{figure}

Next, we further investigated the robustness of these sorting methods against the illumination noise. In mathematical model of the SPI, all noise involved can be treated as the additive noise imposed on the single-pixel measured values. Here, we need to define a signal-to-noise ratio (SNR) of the illumination light field as $\textrm{SNR}_\textrm{I}=10\log_{10}\frac{\left\langle I(c,d)\right\rangle}{\textrm{Std}(\textrm{noise})}$ \cite{YuOR2019,YuCC2019}, where $\langle I(c,d)\rangle=\frac{1}{pq}\sum_{c,d=1}^{p,q}I(c,d)$ denotes the ensemble average, $I(c,d)$ presents one pixel value of the illumination light field, and $\textrm{Std}$ is short for the standard deviation. We chose two distributions of noise for investigation, i.e., normal and Poisson distributions, and used a parrot image as the original image. As mentioned above, the NA and SE orderings are actually not suitable for 2D image reconstruction, thus they will not be compared in the following discussion and analysis. As shown in Fig.~\ref{fig7}, it can be seen that as the $\textrm{SNR}_\textrm{I}$ value increases, the image qualities of the RD, OR, CC and WH orderings all get better. In present of both normal and Poisson distributed noise and with the same $\textrm{SNR}_\textrm{I}$ value, our WH ordering performs slightly better than the CC ordering, and both are significantly better than the RD and OR orderings. The above is enough to demonstrate that our WH method can suppress noise while preserving image details as much as possible.

\subsection{Experimental setup and results}
\begin{figure}[htbp]
\centering\includegraphics[width=\textwidth]{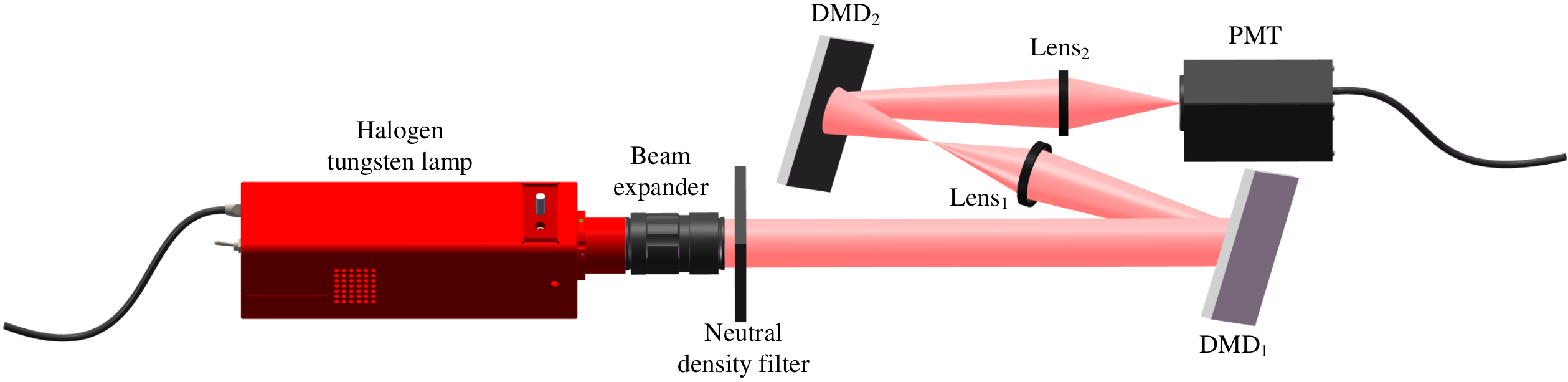}
\caption{Schematic of the experimental setup. A beam of thermal light is expanded, collimated and attenuated, then is reflected by the first digital micromirror device (DMD) onto the plane of the second DMD. The modulated reflection field of the $\textrm{DMD}_2$ (encoded with a series of rearranged Hadamard basis patterns) is collected by a photomultiplier tube (PMT), which works as a single-pixel detector. The reflective object to be sampled is displayed on the $\textrm{DMD}_1$.}
\label{fig8}
\end{figure}

We built an SPI-based experimental setup to further verify the practical imaging performance of this proposed WH sorting method, as shown in Fig.~\ref{fig8}.

The thermal light emitted from a stabilized halogen tungsten lamp (Thorlabs SLS201L/M) was collimated by a beam expander, and then was attenuated to the ultra-weak light level via a series of 2~inch $\times$ 2~inch absorptive neutral density filters (NDFs). Here, we used a 0.7~inch digital micromirror device (DMD, indicated as $\textrm{DMD}_1$) to display the reflective object scene to be imaged (this is a common practice in SPI \cite{Boyd2013,WLiu2019,Yu2021}). The ultra-weak light illuminates the scene, then the reflected light of the latter is incident vertically on and imaged onto the working plane of the second DMD (indicated as $\textrm{DMD}_2$) with $\textrm{lens}_1$, which is encoded with a series of binary modulation patterns. We make the two DMDs parallel to each other to reduce distortion and affine transformations. The core display element of the DMD is a micromirror array sized $768\times1024$, where each micromirror can rotate about a hinge and be individually controlled by the binary value on the corresponding pixel position of a modulated pattern (matrix) to orientate two directions ($\pm12^\circ$ with respect to the normal of the working plane). A deflection of $12^\circ$ (``on'' state) corresponds to entry ``1'' in the pattern, and that of $-12^\circ$ (``off'' state) corresponds to element ``0'' in the pattern. We set the focusing $\textrm{lens}_2$ along the reflection direction of $12^\circ$ micromirror deflection, thus the light falling on the ``0'' pixel position of the pattern will not be collected and appear as dark pixels. We place a counter-type photomultiplier tube (PMT, Hamamatsu H10682-210) on the focal plane of this lens to record the total photon counts and to act as a single-pixel detector. On $\textrm{DMD}_1$, the central $768\times768$ pixels are used to display the object. Since the entries of the Hadamard basis pattern are either 1 or -1 while the DMD can only be loaded with 0-1 matrices, we need to split each Hadamard basis pattern $P_i$ into a pair of complementary 0-1 patterns $P_i^+=(1+P_i)/2$ and $P_i^-=(1-P_i)/2$ so that $P_i=P_i^+-P_i^-$. The complementary patterns in a pair are loaded adjacently onto the $\textrm{DMD}_2$ to realize positive-negative complementary modulation. Then, we also need to accordingly make a difference between every two adjacent photon counts. This strategy is named as complementary differential measurement/modulation \cite{Yu2014,Yu2015,Yu2016,YXLi2019}, and can effectively improve the detection SNR (DSNR). The DSNR can be defined as the power ratio of the signal to the measurement noise, i.e., $\textrm{DSNR}=10\log_{10}Var(y_{signal})/Var(y_{noise})$ \cite{Zhou2019}, where $Var(y_{signal})$ and $Var(y_{noise})$ denote the variances of the single-pixel measured values and measurement noise, respectively.

\begin{figure}[htbp]
\centering\includegraphics[width=\textwidth]{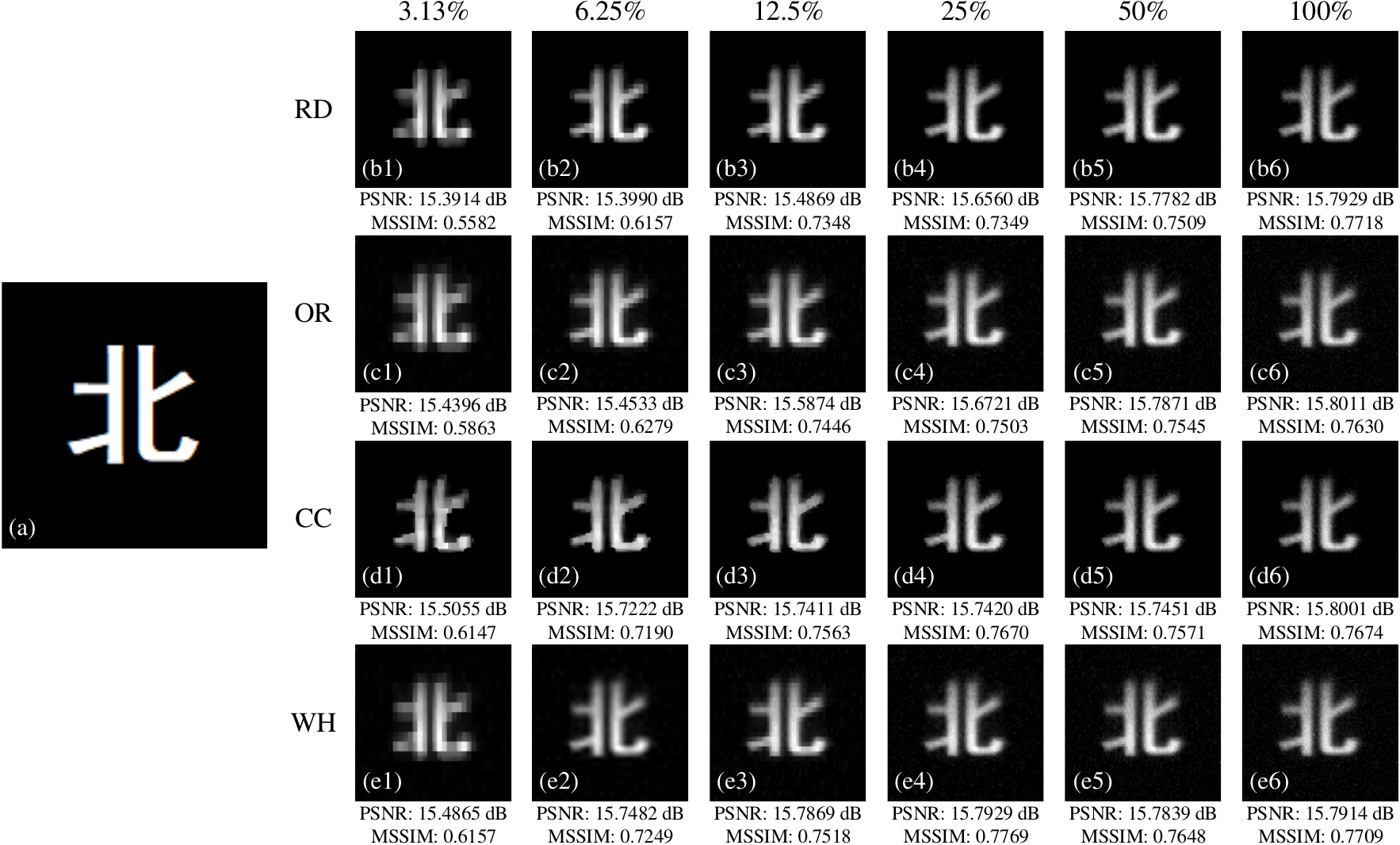}
\caption{Experimental results of a binary object. (a) is a Chinese character ``north'' treated as the reflective binary object to be detected. (b1)--(b6), (c1)--(c6), (d1)--(d6) and (e1)--(e6) give the images of $128\times128$ pixels recovered by using the RD-, OR-, CC- and WH-ordered Hadamard basis patterns under different sampling ratios, respectively. The corresponding PSNRs and MSSIMs are marked right below the figures.}
\label{fig9}
\end{figure}

In experiment, we first used a Chinese character ``north'' as the object scene to be sampled and set the imaging region on the $\textrm{DMD}_2$ to be $128\times128$ pixel-units, each of which occupied $6\times6$ micromirrors, i.e., a total of $768\times768$ micromirrors on the $\textrm{DMD}_2$ were involved in the complementary modulation. We separately loaded the RD-, OR-, CC- and WH-ordered Hadamard basis patterns onto the $\textrm{DMD}_2$, and reconstructed the images from differential single-pixel measurements. The corresponding experimental results were presented in Fig.~\ref{fig9}, with the PSNR and MSSIM values being marked below the recovered images. It can be clearly seen that the PSNRs and MSSIMs of the WH sorting method are higher than those of the RD and OR methods under any sampling rate, and comparable to those of the CC approach but with slightly better visibility. These results are consistent with the simulation results.

\begin{figure}[htbp]
\centering\includegraphics[width=\textwidth]{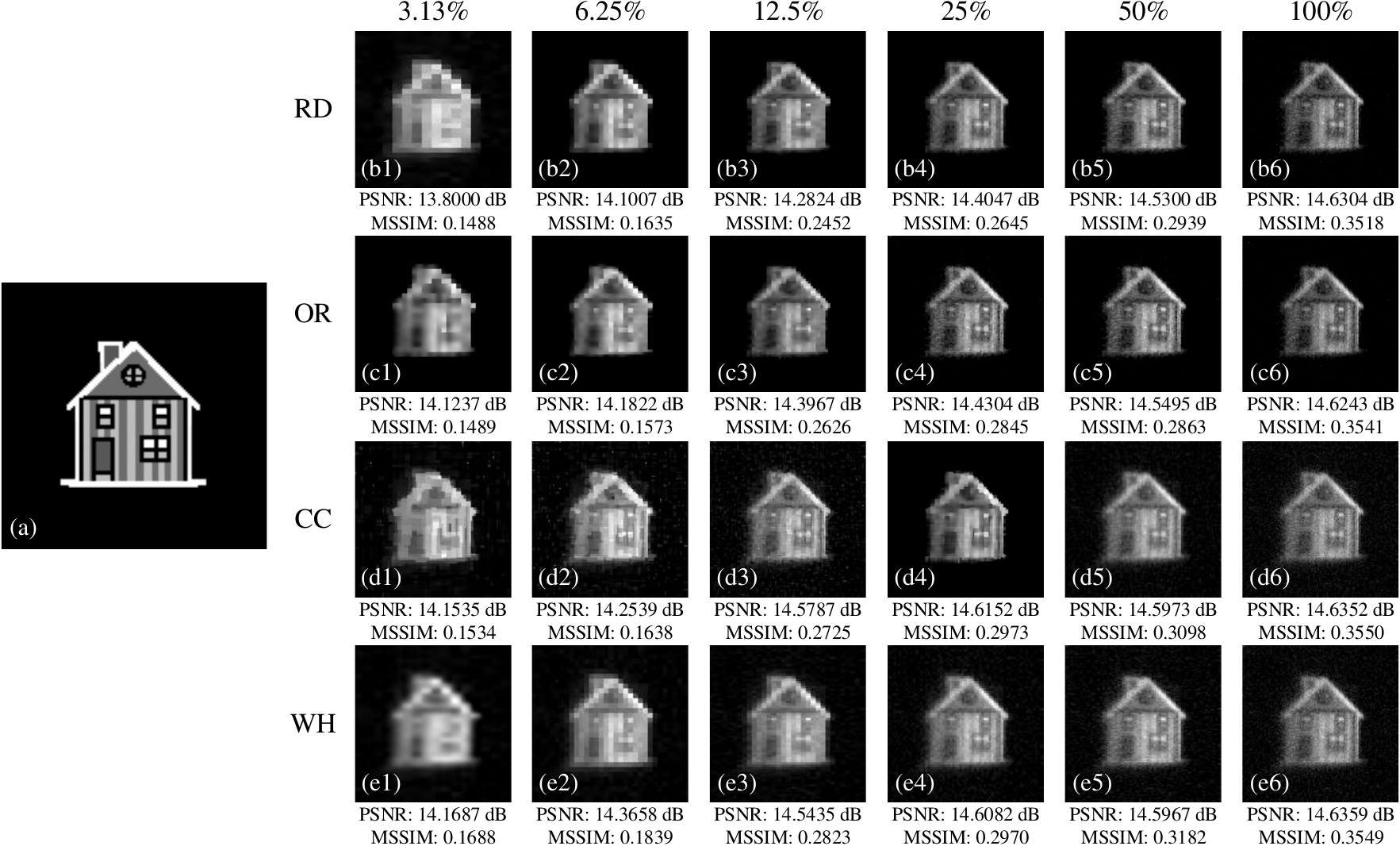}
\caption{Image reconstruction for another grayscale object ``house''. (a) give the object image ``house'' with the grayscale values ranging from 1 to 36; (b1)--(b6), (c1)--(c6), (d1)--(d6) and (e1)--(e6) provide the experimental results retrieved from the RD-, OR-, CC- and WH- ordered Hadamard basis patterns by using different sampling ratios, respectively. All images are of $128\times128$ pixels.}
\label{fig10}
\end{figure}

\begin{figure}[htbp]
\centering\includegraphics[width=\textwidth]{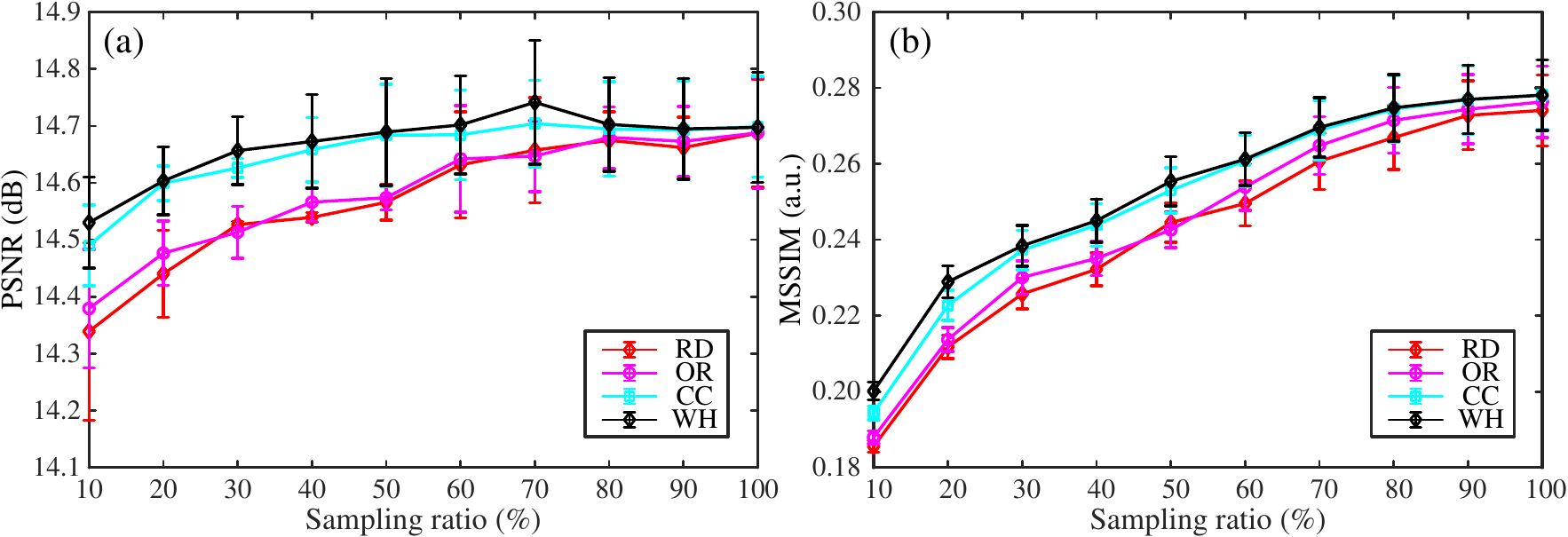}
\caption{Curves of (a) PSNR and (b) MSSIM as a function of the sampling ratio, by applying the RD, OR, CC and WH sorts of the Hadamard basis patterns. The original grayscale images used here for this comparison were ``house'', ``elk'' and ``rocket''. In the curves, the half-height of each error bar denotes the standard deviation of each point.}
\label{fig11}
\end{figure}

\begin{figure}[b!]
\centering\includegraphics[width=\textwidth]{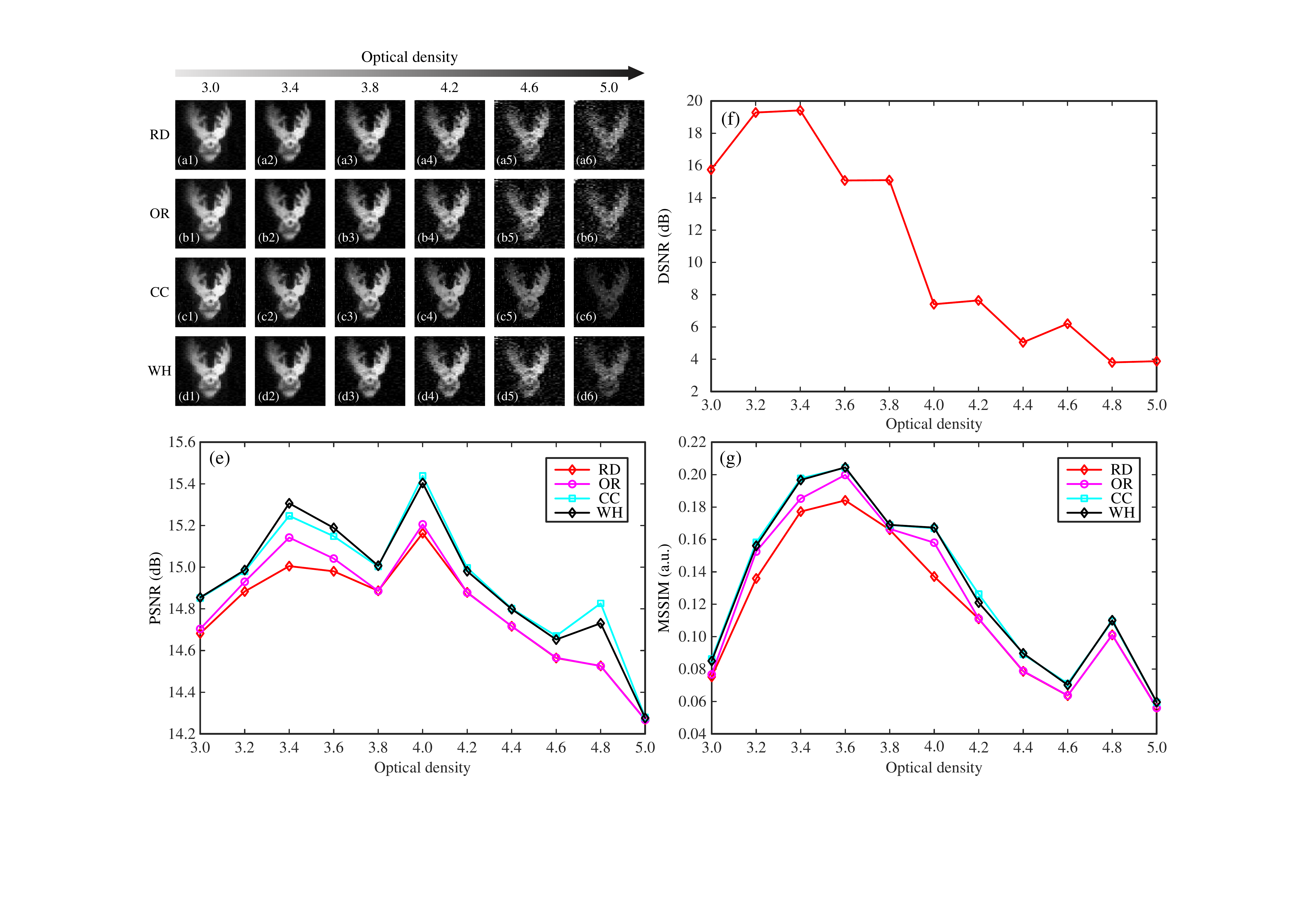}
\caption{Image reconstruction and imaging performance curves under the same sampling ratio of 12.5\% but with different light attenuation of neutral density filters. (a1)--(a6), (b1)--(b6), (c1)--(c6) and (d1)--(d6) give the images recovered by RD, OR, CC and WH methods with neutral density filters of different optical density. (f) shows the relationship between optical densities and detection signal-to-noise ratios. (e) and (f) draw the PSNR and MSSIM curves of these four sorting methods as a function of the optical density of neutral density filters.}
\label{fig12}
\end{figure}

Next, we also tested these orderings with a complex grayscale object ``house'' (see Fig.~\ref{fig10}(a)). As mentioned above, the DMD can only identify 0-1 binary matrix, to display grayscale objects on the $\textrm{DMD}_1$, one can use the pulse-width modulation (PWM) strategy (the grayscale level is determined by the duration time of pixel ``1'', i.e., the time-pulse width) or Floyd-Steinberg error diffusion dithering \cite{Floyd1976} strategy (the grayscale value is depend on the spatial pixel dithering). As we know, the former needs the DMD to encode with multiple binary matrices for each grayscale image's display, sacrificing the display time, while the latter exchanges the spatial pixels for the grayscale level. Given this, here we chose Floyd-Steinberg error diffusion dithering strategy for grayscale image display. Taking a grayscale image of $128\times128$ pixels as an example, its each grayscale pixel was enlarged to a pixel array consisting of $6\times6=36$ sub-pixels, which would be randomly lit up (set to 1). The number of pixels being ``1'' was exactly the gray value of this pixel-unit, thus the gray value ranged from 0 to 36. For the grayscale object ``house'', we performed a series of optical modulation by using the RD, OR, CC and WH orderings of the Hadamard basis patterns and presented their results in Figs.~\ref{fig10}(b1)--\ref{fig10}(b6), \ref{fig10}(c1)--\ref{fig10}(c6), \ref{fig10}(d1)--\ref{fig10}(d6) and \ref{fig10}(e1)--\ref{fig10}(e6). The experimental results of the grayscale image are similar to those of the binary image. The WH ordering outperforms the OR and RD orderings, and is better than the CC sort at the low sampling ratios. Especially at a sampling rate of 3.13\%, the WH method can recover more image details than other sorting methods. After that, we further used other two grayscale objects ``elk'' and ``rocket'' to test the generality of the proposed scheme. In Figs.~\ref{fig11}(a)--\ref{fig11}(b), we plotted the PSNR and MSSIM curves as a function of the sampling ratio. We can see that the imaging performance of the WH and CC orderings is much better than that of the RD and OR orderings, and the WH ordering performs slightly better than the CC ordering (with almost the same standard deviations of MSSIM values).

\begin{figure}[b!]
\centering\includegraphics[width=\textwidth]{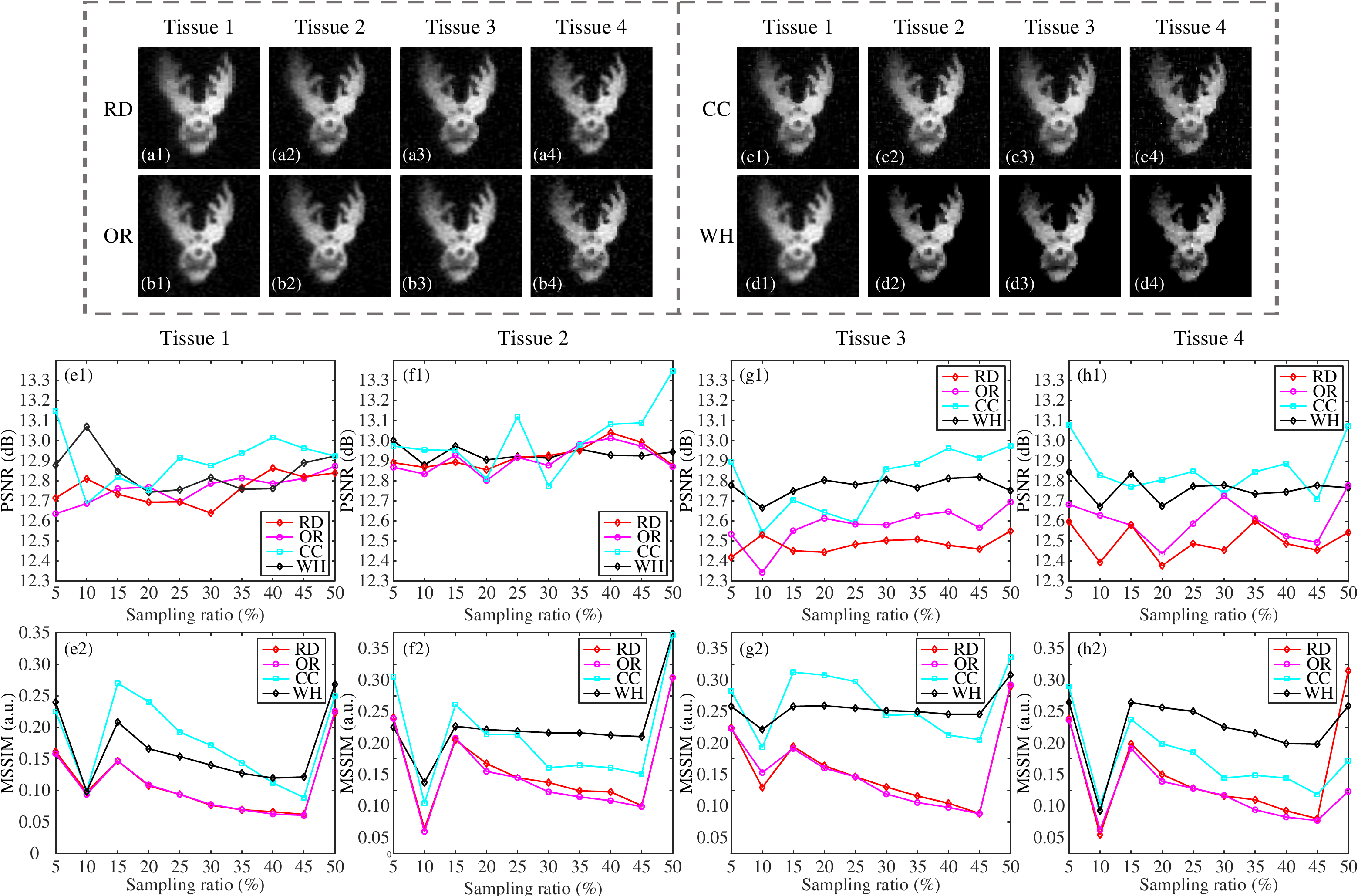}
\caption{Imaging performance in presence of scattering or turbulent medium. (a1)--(a4), (b1)--(b4), (c1)--(c4) and (d1)--(d4) provide the images recovered by RD, OR, CC and WH sorting methods under the same sampling ratio of 12.5\% but with the occlusion of different number of sheets of lens tissue paper before detection. (e1)--(e2), (f1)--(f2), (g1)--(g2) and (h1)--(h2) give the PSNR and MSSIM curves of four orderings with different numbers of sheets of lens tissues (changing from 1 to 4) as a function of the sampling ratio.}
\label{fig13}
\end{figure}

We also investigated the quality of the recovered images by using these four orderings under the same sampling ratio of 12.5\% but with different light attenuation of 2~inch $\times$ 2~inch square absorptive NDFs. Here, we needed to use a metric named optical density (OD) to indicate the attenuation factor, i.e., how much the NDF reduce the optical power of the incident light. It can be defined as $\textrm{OD}=\log_{10}(\frac{1}{\textrm{T}})$, or $\textrm{T}=10^{-\textrm{OD}}$, where $\textrm{T}$ denotes the transmissivity with a value between 0 and 1, and is generally provided in percent (\%). Thus, a higher OD will lead to greater absorption and lower transmission of the incident light. For example, a group consisting of multiple NDFs with a total OD of 3 results in a transmissivity of 0.001, i.e., the light is attenuated to 0.1\% of the incident power. By combining NDFs with different ODs, we made the total OD change from 3.0 to 5.0 in 0.2 intervals. Here, taking the grayscale image ``elk'' as the experimental object, we presented its imaging results in Figs.~\ref{fig12}(a1)--\ref{fig12}(a6), \ref{fig12}(b1)--\ref{fig12}(b6), \ref{fig12}(c1)--\ref{fig12}(c6) and \ref{fig12}(d1)--\ref{fig12}(d6) by using RD, OR, CC and WH methods, respectively, and provided the DSNR values versus the OD values as a reference (see Fig.~\ref{fig12}(f)). When the OD is larger than 3.2, the DSNR decreases with the further increase of the OD. So, to some extent, the increase in OD actually leads to an increase in DSNR. Therefore, the change of DSNR values can be achieved indirectly by switching NDFs of different light attenuation coefficients. Besides, we also drew the PSNR and MSSIM curves as a function of the OD in Figs.~\ref{fig12}(e) and \ref{fig12}(f). It can be seen that when the OD is between 3.4 and 4, the PSNRs and MSSIMs of these four orderings all peak. Therefore, setting OD reasonably can make the DSNR as large as possible. In this experiment, we can see that our WH ordering performs similar to the CC ordering, and much better than RD and OR orderings under ultra-weak light measurement environment.

After that, we placed a few sheets of lens cleaning tissues in front of the PMT to test the imaging ability of these sorting methods in presence of scattering or turbulent medium. The used lens tissues are 4.9~inch $\times$ 2.9~inch organic fiber sheets. When the light passes through lens tissues, some part of the light will be scattered or reflected, some part of the light will be directly transmitted but with the total light intensity being attenuated. In addition, the organic fibers of optical cleaning tissues are chaotically distributed, some areas are relatively agglomerated and some areas are relatively loose. Optically, it is like the compound of multiple point spread functions, which will distort the spatial distribution of the light field, similar to the effect of the turbulent medium. Here, the sampling ratio was also fixed at 12.5\%. By changing the number of sheets of lens tissue paper from 1 to 4, we acquired the corresponding recovered images by using RD, OR, CC and WH methods, as shown in Figs.~\ref{fig13}(a1)--\ref{fig13}(a4), \ref{fig13}(b1)--\ref{fig13}(b4), \ref{fig13}(c1)--\ref{fig13}(c4) and \ref{fig13}(d1)--\ref{fig13}(d4), respectively. Despite adding lens tissues as scattering and turbulent media in the light path, we can still get relatively clear images of the grayscale object. And with the increase in the number of sheets, the image quality will gradually decrease, which is caused by the exacerbation of the combined effect of scattering and point spread functions. Then, we also plotted the PSNR and MSSIM curves of four orderings with different numbers of sheets of lens tissues as a function of the sampling ratio, as shown in Figs.~\ref{fig13}(e1)--\ref{fig13}(e2), \ref{fig13}(f1)--\ref{fig13}(f2), \ref{fig13}(g1)--\ref{fig13}(g2) and \ref{fig13}(h1)--\ref{fig13}(h2), respectively. It can be seen that our WH sorting method performs better (according to MSSIM values) than the CC sorting method when there are multiple sheets of lens tissues as the cover, and much better than the other two orderings under the occlusion of any number of sheets of lens tissues.

\section{Discussion and Conclusion}
In summary, a concept named selection history is proposed to reveal the spatial folding mechanism of the Hadamard matrix, which is widely used for spatial light modulation in the SPI schemes. By recording the Hadamard spatial folding process, a mathematical model is built to derive classic NA and SE orders of the Hadamard basis, as well as the RD, OR and CC orders that have been very popular in recent years. There is evidence that sorting the Hadamard basis is very important for super subsampling and high-quality image reconstruction. Therefore, it is also very crucial to conduct mathematical induction on these orders to deduce the consistent general rules for Hadamard basis sorting and deterministic pattern construction. By using the proposed model, we can easily calculate the number of 1D or 2D connected domains in the Hadamard basis patterns, realize fast conversion between selection history and serial number, and achieve fast mutual conversion between different optimized orderings of the Hadamard basis. We find that the SE ordering is the result of sorting the number of 1D connected domains in patterns while the CC ordering is an increasing order of the number of 2D connected domains. Both RD and OR orderings rely on the grouping and the sort based on 2D connected domains, but the OR order generally performs better than the RD order because it turns coarse grouping into fine grouping. On the basis of this model, we further propose to assign relatively larger weights to the selection positions of the selection history that have a relatively greater impact on the number of 2D connected domains. By this means, the effect of the spatial folding is multiplied. We call it the WH sort of the Hadamard basis, which fully considers the contribution of the folding order to image reconstruction. Both numerical simulation and experimental results have demonstrated that the imaging quality and visibility of the WH order is slightly better than that of the CC order, especially under low sampling ratios, ultra-weak light measurement environment, and in presence of scattering or turbulent medium. In addition, since various optimized orders can be quickly obtained from the selection history, and the multiplication operation on the Hadamard basis can be easily converted into addition and subtraction for fast calculation, once the serial number is known \cite{YuCC2019}. Fast image reconstruction can be achieved without the need to store a large-scale measurement matrix. Therefore, this selection history-based sorting method is very important for the SPI schemes, and we have reasons to believe that this technology may pave the way for real-time high-quality SPI.

\section*{\textsf{Funding}}
Beijing Natural Science Foundation (4222016); Civil Space Project of China (D040301); Youth Talent Promotion Project of the Beijing Association for Science and Technology (none).

\section*{\textsf{Acknowledgment}}
We thank Yu-Xuan Pang for helpful discussions and initial theoretical derivation.

\section*{\textsf{Disclosures}}
The authors declare that there are no conflicts of interest related to this article.

\bigskip





\end{document}